\documentclass[twocolumn,prb,showpacs]{revtex4}
\usepackage{graphicx}
\usepackage{amsmath}
\usepackage{amssymb}
\usepackage{bm}
\usepackage[utf8]{inputenc}

\newcommand{\dr}{\mathrm d \ve r}
\newcommand{\idr}{\, \dr}
\newcommand{\psinket}{\ket{\psi_n}}
\newcommand{\phiaiket}{\ket{\phi_i^a}}
\newcommand{\phiaitildeket}{\ket{\ps \phi_i^a}}
\newcommand{\piatilde}{\ps p_i^a}
\newcommand{\psintildeket}{\ket{\ps \psi_n}}
\newcommand{\psintilde}{\ps \psi_n}
\newcommand{\psinr}{\psi_n(\mathbf{r})}
\newcommand{\psinrtilde}{\ps \psi_n(\mathbf{r})}

\newcommand{\ve}[1]{\bm{\mathrm{#1}}}
\newcommand{\rr}{\ve r}
\newcommand{\bra}[1]{\langle #1 |}
\newcommand{\ket}[1]{| #1 \rangle}
\newcommand{\braket}[1]{\langle #1 \rangle}
\newcommand{\ps}{\tilde}

\newcommand{\dee}{\mathrm d}

\newcommand{\diff}[2]{\frac{\dee #1}{\dee #2}}

\newcommand{\pdiff}[2]{\frac{\partial #1}{\partial #2}}

\newcommand{\fdiff}[2]{\frac{\delta #1}{\delta #2}}

\newcommand{\mrm}{\mathrm}
\newcommand{\nn}{\nonumber}

\newcommand{\pawT}{\mathcal{T}}

\begin{document}
\title{Localized atomic basis set in the projector augmented wave method}
\author{A. H. Larsen}
\affiliation{Center for Atomic-scale Materials Design, Department of
Physics \\ Technical University of Denmark, DK - 2800 Kgs. Lyngby, Denmark}
\author{M. Vanin}
\affiliation{Center for Atomic-scale Materials Design, Department of
Physics \\ Technical University of Denmark, DK - 2800 Kgs. Lyngby, Denmark}
\author{J. J. Mortensen}
\affiliation{Center for Atomic-scale Materials Design, Department of
Physics \\ Technical University of Denmark, DK - 2800 Kgs. Lyngby, Denmark}
\author{K. S. Thygesen}
\affiliation{Center for Atomic-scale Materials Design, Department of
Physics \\ Technical University of Denmark, DK - 2800 Kgs. Lyngby, Denmark}
\author{K. W. Jacobsen}
\affiliation{Center for Atomic-scale Materials Design, Department of
Physics \\ Technical University of Denmark, DK - 2800 Kgs. Lyngby, Denmark}

\date{\today}


%
%
%

\begin{abstract}
  We present an implementation of localized atomic orbital basis sets
  in the projector augmented wave (PAW) formalism within the density
  functional theory (DFT).  The implementation in the real-space GPAW
  code provides a complementary basis set to the accurate but
  computationally more demanding grid representation. The possibility
  to switch seamlessly between the two representations implies that
  simulations employing the local basis can be fine tuned at the end
  of the calculation by switching to the grid, thereby combining the
  strength of the two representations for optimal performance. The
  implementation is tested by calculating atomization energies and
  equilibrium bulk properties of a variety of molecules and solids,
  comparing to the grid results. Finally, it is demonstrated how a grid-quality structure optimization can be performed with significantly reduced
  computational effort by switching between the grid and basis representations.
\end{abstract}

\maketitle
\section{Introduction}\label{sec:intro}
Density functional theory (DFT) with the single-particle
Kohn-Sham (KS) scheme is presently the most widely used method for
electronic structure calculations in both solid state physics and
quantum chemistry.\cite{hohenbergkohn,kohnsham,payne_RMP} Its success is mainly
due to a unique balance between accuracy and efficiency which makes it
possible to handle systems containing hundreds of atoms on a single
CPU with almost chemical accuracy.

At the fundamental level the only approximation of DFT is the
exchange-correlation functional which contains the non-trivial parts
of the kinetic and electron-electron interaction energies. However,
given an exchange-correlation functional one is still left with the
non-trivial numerical task of solving the Kohn-Sham equations. The main
challenge comes from the very rapid oscillations of the valence
electrons in the vicinity of the atom cores that makes it very costly to
represent this part of the wavefunctions numerically. In most modern
DFT codes the problem is circumvented by the use of pseudopotentials.\cite{kleinmanbylander82,vanderbilt,hgh}
The pseudopotential approximation is in principle uncontrolled and is
in general subject to transferability errors. An alternative method is
the projector augmented wave (PAW) method invented by
Blöchl\cite{Blochl:1994}. An appealing feature of the PAW method is
that it becomes exact if sufficiently many projector
functions are used. In another limit the PAW method becomes equivalent
to the ultra-soft pseudopotentials introduced by
Vanderbilt\cite{vanderbilt}.

The representation of the Kohn-Sham wavefunctions is a central aspect
of the numerics of DFT.  High accuracy is achieved by using system
independent basis sets such as plane waves\cite{Blochl:1994,vasp,dacapo},
wavelets\cite{goedecker,arias} or real-space
grids\cite{bernholc96,Mortensen:2005}, which can be systematically
expanded to achieve convergence.  Less accurate but computationally
more manageable methods expand the wavefunction in terms of a
system-dependent localized basis consisting of e.g.
Gaussians\cite{gaussian} or numerical atomic orbitals\cite{Sankey:1989,siesta}.
Such basis sets cannot be systematically enlarged in a simple way, and
consequently any calculated quantity will be subject to basis set
errors. For this reason the former methods are often used to
obtain binding energies where accuracy is crucial, while the latter
are useful for structural properties which are typically less
sensitive to the quality of the wavefunctions.

In this paper we discuss the implementation of a localized atomic
basis set in the PAW formalism and present results for molecular
atomization energies, bulk properties, and structural relaxations. The
localized basis set, which we shall refer to as the LCAO basis, is similar to that of the well-known Siesta pseudopotential 
code\cite{siesta}, but here it is implemented in our recently
developed multigrid PAW code GPAW\cite{Mortensen:2005}. A
unique feature of the resulting scheme is the possibility of using two
different but complementary basis sets: On the one hand wavefunctions
can be represented on a real-space grid which in principle facilitates
an exact representation, and on the other hand the wavefunctions can
be represented in the efficient LCAO basis. This allows the
user to switch seamlessly between the two representations at any point
of a calculation. As a particularly
powerful application of this ``double-basis'' feature, we demonstrate
how accurate structural relaxations can be performed by first relaxing
with the atomic basis set and then switching to the grid for the last
part.  Also adsorption energies, which are typically not very good in
LCAO, can be obtained on the grid at the end of a relaxation.

\begin{figure}
  \includegraphics[width=.42\textwidth]{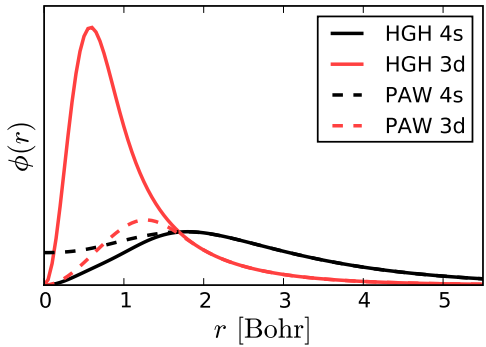}
  \caption{(Color online) The pseudo valence states of iron calculated with PAW and
    the norm-conserving HGH pseudopotentials.  Both methods produce
    smooth wave functions for the delocalized 4s state, but the lack
    of norm conservation allows the short-ranged 3d state in PAW to be
    accurately sampled on a much coarser grid.}
  \label{hghpawcomparison}
\end{figure}

While LCAO pseudopotential codes as well as plane-wave/grid PAW codes
already exist and have been discussed extensively in the
literature,\cite{Sankey:1989,siesta,Blochl:1994} the combination of
LCAO and PAW is new. Compared to the popular Siesta method, which is
based on norm-conserving pseudopotentials, the advantage of the
present scheme (apart from the ``double basis'' feature) is that PAW
works with coarser grids to represent the density and effective
potentials.  As an example, Fig.\ \ref{hghpawcomparison} shows the
atomic orbitals of iron calculated with the norm-conserving
Hartwigsen-Goedecker-Hutter (HGH) pseudopotentials\cite{hgh} as well
as with PAW. Clearly the $d$ wavefunction is much smoother in
PAW.  This is essential for larger systems where operations on the
grid, i.e.\ solving the Poisson equation, evaluating the
density, and calculating the potential matrix elements, become
computationally demanding.

\section{Projector Augmented Wave Method}
In this section we give a brief review of the PAW formalism. For
simplicity we restrict the equations to the case of spin-paired, finite
systems, but the generalizations to magnetic and periodic systems are 
straightforward. For a more comprehensive presentation we refer to Ref.\ \onlinecite{Blochl:1994}.

\subsection{PAW Transformation Operator}
The PAW method is based on a linear transformation $\pawT$ which maps
some computationally convenient ``pseudo'' or ``smooth'' wavefunctions
$\ket{\ps \psi_n}$ to the physically relevant ``all-electron''
wavefunctions $\ket{\psi_n}$:
\begin{align}
  \label{eq:transformation}
  \psinket = \pawT \psintildeket,
\end{align}
where $n$ is a quantum state label, consisting of a band index and
possibly a spin and $\ve k$-vector index.

The transformation is chosen as $\pawT = 1 + \sum_a \pawT^a$, i.e.\
the identity operator plus an additive contribution centered around
each atom, which differs based on the species of atom.

The atomic contribution for atom $a$ is determined by choosing a set
of smooth functions $\ps \phi_i^a(\ve r)$, called pseudo partial
waves, and requiring the transformation to map those onto the atomic
valence orbitals $\phi_i^a(\ve r)$ of that atom, called all-electron
partial waves.  This effectively allows the all-electron behaviour to
be incorporated by the smooth pseudo wave functions.  Since the
all-electron wave functions are smooth sufficiently far from the atoms,
we may require the pseudo partial waves to match the all-electron ones
outside a certain cutoff radius, such that $\ps \phi_i^a(\ve r) =
\phi_i^a(\ve r)$ for $r > r_c$.  This localizes the atomic
contribution $\pawT^a$ to the \emph{augmentation sphere} $r < r_c$.
Finally a set of localized \emph{projectors} $\ps p_i^a(\ve r)$ is
chosen as a dual basis to the pseudo partial waves.  We further want
the partial wave-projector basis to be complete within the
augmentation sphere, in the sense that any pseudo wave function should
be expressible in terms of pseudo partial waves, and therefore require
\begin{align}
  \sum_i \ket{\ps \phi_i^a}\bra{\ps p_i^a} = 1,\qquad
  \braket{\ps \phi_i^a| \ps p_j^a} = \delta_{ij}.\label{completeness}
\end{align}
The transformation $\pawT$ is then defined by
\begin{align}
  \label{eq:transformation1}
  \pawT = 1 + \sum_a \sum_i (\phiaiket - \phiaitildeket)
  \bra{\piatilde},
\end{align}
which allows the all-electron Kohn-Sham wavefunction $\psi_n(\ve
r)=\braket{\ve r|\psi_n}$ to be recovered from a pseudo wave function
through
\begin{align}
  \label{eq:psin_final}
  \psinr = \psinrtilde + \sum_a \sum_i (\phi_i^a(\mathbf{r}) - \ps
  \phi_i^a(\mathbf{r})) \braket{\piatilde|\psintilde}.
\end{align}
We emphasize that the all-electron wave functions are never evaluated
explicitly, but all-electron values of observables are calculated
through manipulations which rely only on coarse grids or
one-dimensional radial grids.  Using using Eqs.\
\eqref{eq:transformation} and \eqref{eq:transformation1}, the
all-electron expectation value for any semi-local operator $O$ due to
the valence states can be written
\begin{align}
  \braket O &= 
  \sum_n f_n \braket{\ps \psi_n | O | \ps \psi_n} \nn\\
  &\quad+ \sum_{naij} f_n \braket{\ps \psi_n | \ps p_i^a}
  \braket{\phi_i^a | O | \phi_j^a}
  \braket{\ps p_j^a | \ps \psi_n} \nn\\
  &\quad - \sum_{naij} f_n\braket{\ps \psi_n | \ps p_i^a}
  \braket{\ps \phi_i^a | O | \ps \phi_j^a}
  \braket{\ps p_j^a | \ps \psi_n}.
  \label{expectationvalue1}
\end{align}
Inside the augmentation spheres the partial wave expansion is ideally
complete, so the first and third terms will cancel and leave only the
all-electron contribution.  Outside the augmentation spheres the
pseudo partial waves are identical to the all-electron ones, so the
two atomic terms cancel.  The atomic matrix elements of $O$ in the
second and third terms can be pre-evaluated for the isolated atom on
high-resolution radial grids, so operations on smooth quantities, like
$\braket{\ps \psi_n|O|\ps\psi_n}$ and $\braket{\ps p_i^a|\ps \psi_n}$,
are the only ones performed during actual calculations.

It is convenient to define the \emph{atomic density matrices}
\begin{align}
  D_{ij}^a = \sum_n \braket{\ps p_i^a | \ps \psi_n}
  f_n \braket{\ps \psi_n | \ps p_j^a},
  \label{pawatomicdensitymatrices}
\end{align}
since these completely describe the dependence of the atomic terms in
Eq.\ \eqref{expectationvalue1} on the pseudo wave functions.  The
expectation value can then be written
\begin{align}
  \braket{O} &= \sum_n f_n \braket{\ps \psi_n|O|\ps \psi_n}\nn\\
  &\quad+ \sum_{aij} D_{ji}^a \left(\braket{\phi_i^a|O|\phi_j^a} - 
    \braket{\ps \phi_i^a|O|\ps \phi_j^a}
    \right).\label{expectationvalue}
\end{align}
Although the PAW method is an exact implementation of density
functional theory, some approximations are needed for realistic
calculations. The frozen-core approximation assumes that the core
states are localized within the augmentation spheres and that they are
not modified by the chemical environment and hence taken from atomic
reference calculations. The non-completeness of the basis, or
equivalently the finite grid-spacing, will introduce an error in the
evaluation of the PS contribution $\ps \psi_n$ in
\eqref{expectationvalue1}. Finally, the number of partial waves and
projector functions is obviously finite.  This means that the
completeness conditions of Eq.\ \eqref{completeness} we have required 
are not strictly fulfilled.
This approximation can be controlled directly by
increasing the number of partial waves and projectors.

\subsection{Density}
The electron density $n(\ve r)$ is the expectation value
of the real-space projection operator and, by Eq.\
\eqref{expectationvalue}, takes the form
\begin{align}
  \label{eq:paw-density}
  n(\ve r) = \ps n(\ve r) 
+ \sum_a \left [n^a(\ve r - \ve R^a) - \ps n^a(\ve r - \ve R^a)\right],
\end{align}
where
\begin{align}
  \label{eq:atomic_densities}
  \ps n(\ve r) &= \sum_n f_n |\psinrtilde|^2 
  + \sum_a \ps n_c^a(|\ve r - \ve R^a|), \\
  n^a(\rr) &= \sum_{ij} D_{ji}^a \phi_i^a(\rr) \phi_j^a(\rr) + n_c^a(r), \\
  \ps n^a(\rr) &= \sum_{ij} D_{ji}^a \ps \phi_i^a(\rr) \ps \phi_j^a(\rr) +
  \ps n_c^a(r).
\end{align}
Here we have separated out the all-electron core density $n_c^a(r)$ 
and the pseudo core density $\ps n_c^a(r)$, where the latter
can be chosen as any smooth continuation of $n_c^a(r)$ inside the
augmentation spheres, since it will cancel out in Eq.\ \eqref{eq:paw-density}.
We omit conjugation of the 
partial waves since
these can be chosen as real functions without loss of generality.

\subsection{Compensation charges}
In order to avoid dealing with the cumbersome nuclear point charges,
and to compensate for the lack of norm-conservation, we introduce
smooth localized \emph{compensation charges} $\ps Z^a(\ve r)$ on each
atom, which are added to $\ps n(\ve r)$ and $\ps n^a(\ve r)$, thus
keeping the total charge neutral.  This yields a total charge
density that can be expressed as
\begin{align}
\rho(\ve r) = \ps \rho(\ve r) 
+ \sum_a\left[\rho^a(\ve r - \ve R^a) - \ps \rho^a(\ve r- \ve R^a)\right],
\end{align}
in terms of the neutral charge densities
\begin{align}
  \ps \rho(\ve r) &= \ps n(\ve r) + \ps Z(\ve r) 
  = \ps n(\ve r) + \sum_a \ps Z^a(\ve r - \ve R^a),\\
  \rho^a(\ve r) &= n^a(\ve r) + \mathcal Z^a \delta(\ve r),\\
  \ps \rho^a(\ve r) &= \ps n^a(\ve r) + \ps Z^a(\ve r),
\end{align}
where $\mathcal Z^a \delta(\ve r)$ is the central nuclear point
charge.  The compensation charges are chosen to be
localized functions around each atom of the form
\begin{align}
  \ps Z^a(\ve r) = \sum_L Q_L^a \ps g_L^a(\ve r)
  = \sum_{lm} Q_{lm}^a r^l \ps g_l^a(r) Y_{lm}(\hat{\ve r}),
\end{align}
where $\ps g_l^a(r)$ are fixed Gaussians, and $Y_{lm}(\hat{\ve r})$
are spherical harmonics.  We use $L=l,m$ as a composite index for
angular and magnetic quantum numbers.  The expansion coefficients
$Q_L^a$ are determined in terms of $D_{ij}^a$ by requiring the
compensation charges to cancel all the multipole moments of each
augmentation region up to some order, 
generally $l_{\mrm{max}}=2$.  The charges will therefore
dynamically adapt to the surroundings of the atom.
For more details we refer to the original work by
Blöchl\cite{Blochl:1994}.

\subsection{Total Energy}
The total energy can also be separated into smooth and atom-centered
contributions
\begin{align}
  E = \ps E + \sum_a (E^a - \ps E^a),
  \label{totalenergy}
\end{align}
where
\begin{align}
  \ps E &=
  \sum_n f_n \braket{\psintilde|-\tfrac{1}{2} \nabla^2|\psintilde}
  + \sum_a \int \ps n(\ve r) \bar v^a(|\ve r - \ve R^a|) \idr \nn\\
  &\quad+ \frac{1}{2} \iint \frac{\ps \rho(\mathbf{r})
    \ps \rho(\mathbf{r'})}{|\mathbf{r} - \mathbf{r'}|} 
  \idr \idr'
  + E_{xc}[\ps n], \label{etilde}\\
  E^a &= \sum_{ij} D_{ji}^a \braket{\phi^{a}_i|-\tfrac{1}{2}
    \nabla^2|\phi^{a}_j} + T_{\mrm{core}}^a
  \nn\\
  &+ \frac12 \iint \frac{\rho^a(\ve r) \rho^a(\ve r')}{|\ve r - \ve r'|}
  \idr \idr' + E_{xc}[n^a], \\
  \ps E^a &= \sum_{ij} D_{ji}^a \braket{\ps \phi^a_i|-\tfrac{1}{2}
    \nabla^2|\ps \phi^a_j} 
  + \ps T_{\mrm{core}}^a + \int \ps n^a(\ve r) \bar v^a(r) \idr\nn\\
  &+ \frac12 \iint \frac{\ps \rho^a(\mathbf{r})
    \ps \rho^a(\mathbf{r'})}{|\mathbf{r} - \mathbf{r'}|} \idr \idr'
  + E_{xc}[\ps n^a].
\end{align}
The terms $T_{\mrm{core}}^a$ and $\ps T_{\mrm{core}}^a$ are the
kinetic energy contributions from the frozen core states, while $\bar
v^a(r)$ is an arbitrary potential, vanishing for $r>r^a_c$.  This
potential is generally chosen to make the atomic potential smooth,
while its contribution to the total energy vanishes if the partial
wave expansion is complete.\cite{Mortensen:2005}

$E_{\mrm{xc}}$ is the exchange-correlation functional, which must be
local or semilocal as per Eq.\ \eqref{expectationvalue} for the above
expressions to be correct.  While the functional is non-linear, it
remains true that
\begin{align}
  E_{\mrm{xc}}[n]
  = E_{\mrm{xc}}[\ps n] 
  + \sum_a\left(
    E_{\mrm{xc}}[n^a]  - E_{\mrm{xc}}[\ps n^a]
    \right),
\end{align}
because of the functional's semilocality: the energy contribution from
$\ps n(\ve r)$ around every point inside the augmentation sphere is
exactly cancelled by that of $\ps n^a(\ve r)$, since $\ps n(\ve r)$
and $\ps n^a(\ve r)$ are exactly identical here, leaving only the
contribution $E_{\mrm{xc}}[n^a]$.  Outside the augmentation region, a
similar argument applies to $n^a(\ve r)$ and $\ps n(\ve r)$, leaving
only the energy contribution from $\ps n(\ve r)$ which is here equal
to the all-electron density.

\subsection{Hamiltonian and orthogonality}
In generic operator form, the Hamiltonian corresponding to the total
energy from Eq.\ \eqref{totalenergy} is
\begin{align}
  \label{eq:paw_hamiltonian}
  \ps H = -\tfrac12 \nabla^2 + \ps v
  + \sum_{aij} \ket{\ps p_i^a} \Delta H_{ij}^a \bra{p_j^a},
\end{align}
where $\ps v = \ps v_{\mrm{Ha}}[\ps \rho] + \bar v + v_{\mrm{xc}}[\ps
n]$ is the local effective potential, containing the Hartree, the
arbitrary localized and the xc potentials, and where
\begin{align}
  \Delta H_{ij}^a = \pdiff{E}{D_{ji}^a} \label{DeltaH}
\end{align}
are the \emph{atomic Hamiltonians} containing the atom-centered
contributions from the augmentation spheres.  Since the all-electron
wave functions $\psi_n$ must be orthonormal, the pseudo wave functions
$\ps \psi_n$ must obey
\begin{align}
  \delta_{nm} = \braket{\psi_n | \psi_m}
  = \braket{\ps \psi_n | \pawT^\dagger \pawT | \ps \psi_m}
  = \braket{\ps \psi_n | S | \ps \psi_m},
\end{align}
where we have defined the \emph{overlap operator}
\begin{align}
  S = \pawT^\dagger \pawT 
=1 + \sum_{aij} \ket{\ps p_i^a} \Delta S_{ij}^a \bra{\ps p_j^a}.
  \label{overlapoperator}
\end{align}
The atomic contributions
\begin{align}
  \Delta S_{ij}^a= \braket{\phi_i^a|\phi_j^a} - \braket{\ps\phi_i^a|\ps\phi_j^a} 
\end{align}
are constant for a given element.

Given the Hamiltonian and orthogonality condition, a variational
problem can be derived for the pseudo wave functions.  This problem is
equivalent to the generalized Kohn-Sham eigenvalue problem
\begin{align}
  \ps H \ket{\ps \psi_n} = S \ket{\ps \psi_n} \epsilon_n,
\label{paweigenvalueproblem}
\end{align}
which can then be solved self-consistently with available techniques.

\section{Localized basis sets in PAW} 
We now introduce a set of
basis functions $\ket{\Phi_\mu}$ which are fixed, strictly localized atomic
orbital-like functions represented numerically, following the approach by
Sankey and Niklewski\cite{Sankey:1989}. We furthermore consider the pseudo
wave functions $\ket{\ps \psi_n}$ to be linear combinations of the new 
basis functions
\begin{align}
  \label{eq:basis}
  \ket{\ps \psi_n} 
  = \sum_\mu c_{\mu n} \ket{\Phi_\mu},
\end{align}
where the coefficients $c_{\mu n}$ are variational parameters.
It proves useful to define the density matrix
\begin{align}
  \rho_{\mu\nu} = \sum_n c_{\mu n} f_n c_{\nu n}^*.\label{densitymatrix}
\end{align}
The pseudo density can be evaluated from the density matrix through
\begin{align}
  \ps n(\ve r) = \sum_{\mu\nu} \Phi_\mu^*(\ve r)\Phi_\nu(\ve r)
  \rho_{\nu\mu} + \sum_a \ps n_c^a(\ve r).
  \label{lcaopseudodensity}
\end{align}
Ahead of a calculation, we evaluate the matrices
\begin{align}
  T_{\mu\nu} &= \braket{\Phi_\mu | -\tfrac12 \nabla^2 | \Phi_\nu},
\label{matrix_elements0}\\
  P_{i\mu}^a &= \braket{\ps p_i^a | \Phi_\mu},\\
  \Theta_{\mu\nu} &= \braket{\Phi_\mu | \Phi_\nu},
\label{matrix_elements}
\end{align}
which are used to evaluate most of the quantities of the previous sections in
matrix form.
The atomic density matrices from Eq.\ \eqref{pawatomicdensitymatrices} become
\begin{align}
  D_{ij}^a = \sum_{\mu\nu} P_{i\mu}^a \rho_{\mu\nu} P_{j\nu}^{a*},
\end{align}
and the kinetic energy contribution in the first term of Eq.\ \eqref{etilde} is
\begin{align}
  \sum_n f_n \braket{\ps \psi_n|-\tfrac12 \nabla^2|\ps \psi_n}
  = \sum_{\mu\nu} T_{\mu\nu} \rho_{\nu\mu}.
\end{align}
We can then define the Hamiltonian matrix elements by taking the
derivative of the total energy $E$ with respect to the density matrix
elements, which eventually results in the discretized Hamiltonian
\begin{align}
  H_{\mu\nu} \equiv \pdiff{E}{\rho_{\nu\mu}}
  = T_{\mu\nu} + V_{\mu\nu} 
  + \sum_{aij} P_{i\mu}^{a*} \Delta H_{ij}^a P_{j\nu}^a,
  \label{hamiltonianmatrix}
\end{align}
where
\begin{align}
  V_{\mu\nu} = \int \Phi_\mu^*(\ve r) \ps v(\ve r) \Phi_\nu(\ve r) \idr.
  \label{potentialmatrix}
\end{align}
The overlap operator of Eq.\ \eqref{overlapoperator} has the matrix
representation
\begin{align}
  S_{\mu\nu} = \braket{\Phi_\mu|S|\Phi_\nu}=
  \Theta_{\mu\nu} + \sum_{aij} P_{i\mu}^{a*} \Delta S_{ij}^a P_{j\nu}^a,
\end{align}
so orthogonality of the wave functions is now expressed by
\begin{align}
  \sum_{\mu\nu} c_{\mu m}^* S_{\mu \nu} c_{\nu n} = \delta_{mn}.
  \label{orthogonalitycondition}
\end{align}
This is incorporated by defining a quantity $\Omega$ to be
variationally minimized with respect to the coefficients, specifically
\begin{align}
  \Omega = E - \sum_{mn\mu\nu} \lambda_{nm}\left(
    c_{\mu m}^* S_{\mu \nu} c_{\nu n} - \delta_{mn}
  \right).
\end{align}
Setting the derivative of $\Omega$ with respect to $c_{\mu n}$ equal
to 0, one obtains the generalized eigenvalue equation
\begin{align}
  \sum_\nu H_{\mu\nu} c_{\nu n} 
  = \sum_{\nu} S_{\mu\nu} c_{\nu n} \epsilon_n,
  \label{eigenvalueproblem}
\end{align}
which can be solved for the coefficients $c_{\mu n}$ and
energies $\epsilon_n$ when the Hamiltonian $H_{\mu\nu}$ and the
overlap matrix $S_{\mu \nu}$ are known.

\subsection{Basis functions generation}
The basis functions $\ket{\Phi_\mu}$ in Eq.\ \eqref{eq:basis} are 
atom-centered orbitals written as products of numerical radial functions
and spherical harmonics:
\begin{align}
 \Phi_{nlm}(\ve r) = \varphi_{nl}(r) Y_{lm}(\hat{\ve r}).
 \label{basisfunctionform}
\end{align}
In order to make the Hamiltonian and overlap matrices sparse in the
basis-set representation, we use strictly
localized radial functions, i.e.\ orbitals that are identically zero
beyond a given radius, as proposed by Sankey and Niklewski
\cite{Sankey:1989} and successfully implemented in the SIESTA method
\cite{siesta}.

\begin{figure}
  \includegraphics[width=.45\textwidth]{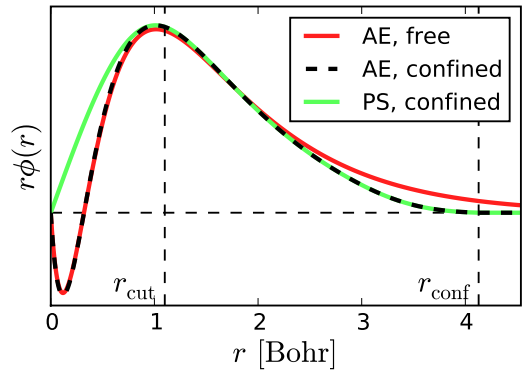}
  \caption{NAO generation for the nitrogen 2s state: the all-electron
    orbital of the free atom, the confined all-electron orbital, and
    the corresponding pseudo wave function after applying the inverse
    PAW transformation.  The augmentation sphere and basis function
    cutoffs are indicated.}
  \label{fig:basisgeneration}
\end{figure}

The first (single-zeta) basis orbitals $\varphi_{nl}^{\mathrm{AE}}(r)$ are
obtained for each valence state by solving the radial all-electron
Kohn-Sham equations for the isolated atom in the presence of a
confining potential with a certain cutoff.  If the confining potential
is chosen to be smooth, the basis functions similarly become smooth.
We use the same confining potential as proposed in Ref.\
\onlinecite{siesta2}.  The smooth basis functions are then obtained
using $\varphi_{nl}(r) = \pawT^{-1} \varphi_{nl}^{\mathrm{AE}}(r)$.
The result of the procedure is illustrated in Fig.\ \ref{fig:basisgeneration}.


The cutoff radius is selected in a systematic way by specifying the
energy shift $\Delta E$ of the confined orbital compared to the
free-atom orbital. In this approach small values of $\Delta E$ will
correspond to long-ranged basis orbitals\cite{siesta}.

To improve the radial flexibility, extra basis functions with the same
angular momentum $l$ (multiple-zeta) are constructed for each valence
state using the split-valence technique\cite{siesta}. The extra
function is constructed by matching a polynomial to the tail of the
atomic orbital, where the matching radius is determined by requiring
the norm of the part of the atomic orbital outside that radius to have
a certain value.

Finally, polarization functions (basis functions with $l$ quantum
number corresponding to the lowest unoccupied angular momentum) can be
added in order to improve the angular flexibility of the basis. There
are several approaches to generate these orbitals, such as perturbing
the occupied eigenstate with the highest $l$ quantum number with an
electric field using first order perturbation theory (like in Ref.\
\onlinecite{siesta}) or using the appropriate unoccupied orbitals.  As
a first implementation we use a Gaussian-like function of the form
$r^l \exp(-\alpha r^2)$ for the radial part, where $l$ corresponds to
the lowest unoccupied angular momentum.  This produces reasonable
polarization functions as demonstrated by the results presented in a
following section. 

A generator program is included in the GPAW code and it can produce
basis sets for virtually any elements in the periodic table.
Through our experiences with generating and using different basis
sets, we have reached the following set of default parameters: We
usually work with a DZP basis. The energy shift for the atomic orbital
is taken as 0.1 eV, and the tail-norm is 0.16 (in agreement with
SIESTA\cite{siesta}).  The width of the Gaussian used for the
polarization function is 1/4 of the cut-off radius of the first zeta
basis function. Further information can be found in the documentation
for the basis set generator.
At this
point we have not yet systematically optimized the basis set
parameters, although we expect to do so by means of an automatic
procedure.

\subsection{Atomic forces}
The force on some atom $a$ is defined as the negative
derivative of the total energy of the system with respect to the
position of that atom,
\begin{align}
  \ve F^a = -\pdiff{E}{\ve R^a}.
\end{align}
The derivative is to be taken with the constraints that
selfconsistency and orthonormality according to
\eqref{orthogonalitycondition} must be obeyed.  This implies that the
calculated force will correspond to the small-displacement limit of
the finite-difference energy gradient one would obtain by performing
two separate energy calculations, where atom $a$ is slightly displaced in
one of them.

The expression for the force is obtained by using the chain rule on the total
energy of Eq.\ \eqref{totalenergy}. The primary complication compared to the
grid-based PAW force formula, Eq.\ (50) from Ref.~\onlinecite{Mortensen:2005},
is that the basis functions move with the atoms, introducing extra terms in
the derivative.

The complete formula for the force on atom $a$ is
\begin{align}
  \ve F^a
  &= 
  2 \Re \sum_{\mu \in a; \nu} \diff{T_{\mu\nu}}{\ve R^a} \rho_{\nu\mu}
  - 2 \Re \sum_{\mu \in a;\nu} \diff{\Theta_{\mu\nu}}{\ve R^a} E_{\nu\mu} 
  \nn\\ \quad&
  + 2 \Re \sum_{b; \mu \in a; \nu} \ve Z_{\mu\nu}^b E_{\nu\mu}
  - 2 \Re \sum_{\mu\nu} \ve Z_{\mu\nu}^a E_{\nu\mu} 
  \nn\\ \quad&
  - 2 \Re \sum_{b; \nu; \mu \in a} \ve A_{\mu\nu}^b \rho_{\nu\mu}
  + 2 \Re \sum_{\mu\nu} \ve A_{\mu\nu}^a \rho_{\nu\mu}
  \nn\\ \quad&
  +2\Re\sum_{\mu \in a; \nu} \left[ \int
    \diff{\Phi_\mu^*(\ve r)}{\ve R^a} \ps v(\ve r) \Phi_\nu(\ve r) \idr
  \right] \rho_{\nu\mu}
  \nn\\ \quad&
  -\int \ps v(\ve r) \diff{\ps n_c^a(|\ve r - \ve R^a|)}{\ve R^a} \idr
  -\int \ps n(\ve r) \diff{\bar v^a(|\ve r - \ve R^a|)}{\ve R^a} \idr
  \nn\\ \quad&
  - \int \ps v_H(\ve r) \sum_L Q_L^a \diff{\ps g_L^a(\ve r - \ve R^a)}{\ve R^a} \idr,
  \label{forces}
\end{align}
where
\begin{align}
  \ve A_{\mu\nu}^b &= 
  \sum_{ij} \diff{P_{i \mu}^{b*}}{\ve R^a} \Delta H_{ij}^b P_{j \nu}^b,\\
  \ve Z_{\mu\nu}^b &= 
  \sum_{ij} \diff{P_{i \mu}^{b*}}{\ve R^a} \Delta S_{ij}^b P_{j \nu}^b,\\
  E_{\mu\nu} &= \sum_{\lambda\xi} S^{-1}_{\mu \lambda} H_{\lambda \xi} \rho_{\xi \nu}.
  \label{energydensitymatrix}
\end{align}
The notation $\mu \in a$ denotes that summation
should be performed only over those basis functions that reside on
atom $a$.  

Eq.\ \eqref{forces} is derived in Appendix \ref{appendixforces}.  The
last three terms are basis set independent, and inherited from the
grid-based implementation.

\section{Implementation}
The LCAO code is implemented in GPAW, a real-space PAW code. For the
details of the real-space implementation we refer to the original
paper\cite{Mortensen:2005}. In this code the density, effective
potential and wave functions are evaluated on real-space grids.

In LCAO the matrix elements of
the kinetic and overlap operators $T_{\mu\nu}$, $\Theta_{\mu\nu}$ and
$P_{i\mu}^a$ in Eqs. \eqref{matrix_elements0}-\eqref{matrix_elements} 
are efficiently calculated in
Fourier space based on analytical expressions\cite{Sankey:1989}.
For each pair of different
basis orbitals (i.e.\ independently of the atomic positions), the
overlap can be represented in the form of radial functions and
spherical harmonics.  These functions are stored as splines which can in turn 
be evaluated for a multitude of different atomic separations.

The two-center integrals are thus calculated once for a given atomic
configuration ahead of the self-consistency loop. This is
equivalent to the SIESTA approach\cite{siesta}.

The matrix elements of the effective potential $V_{\mu\nu}$ are still
calculated numerically on the three dimensional real-space grid, since
the density is also evaluated on this grid\cite{Mortensen:2005}.

Because of the reduced degrees of freedom of a basis calculation
compared to a grid-based calculation, the Hamiltonian from 
Eq.\ \eqref{hamiltonianmatrix} is directly
diagonalized in the space of the basis functions 
according to Eq.\ \eqref{eigenvalueproblem}.  This considerably lowers
the number of required iterations to reach selfconsistency, compared to
the iterative minimization schemes used in grid-based calculations.

For each step in the selfconsistency loop, the Hartree potential $\ps
v_{\mathrm{Ha}}(\ve r)$ is calculated by solving the Poisson equation $\nabla^2
\ps v_{\mathrm{Ha}}(\ve r) = -4\pi \ps \rho(\ve r)$ in real
space using existing multigrid methods, such as the Gauss-Seidel and
Jacobi methods. A solver based on the fast Fourier transform is also
available in the GPAW code.

The calculations are parallelized over k-points, spins and real-space
domains like in the grid-based case\cite{Mortensen:2005}.  We further
distribute the orbital-by-orbital matrices such as $H_{\mu\nu}$ and
$S_{\mu\nu}$, and use ScaLAPACK for operations on these, notably the
diagonalization of Eq.\ \eqref{eigenvalueproblem}.

\subsection{Localized functions on the grid}
Quantities such as the density $\ps n(\ve r)$ and effective potential
$\ps v(\ve r)$ are still stored on 3D grids.  Matrix elements like
$V_{\mu\nu}$ in Eq.\ \eqref{potentialmatrix}, and the pseudo density
given by Eq.\ \eqref{lcaopseudodensity}, can therefore be calculated
by loops over grid points.

Since each basis function is nonzero only in a small part of space, we
only store the values of a given function within its bounding sphere.
Each function value inside the bounding sphere is calculated as the
product of radial and angular parts vz.\ Eq.\ \eqref{basisfunctionform},
where the radial part is represented by a spline, and the spherical
harmonic evaluated in cartesian form, i.e.\ as a polynomial.  The same
method is used to evaluate derivatives in force calculations, although
this involves the derivatives of these quantities aside from just
their function values.

We initially compile a data structure to keep track of which functions
are nonzero for each grid point.  When looping over the grid, we
maintain a list of indices $\mu$ for the currently nonzero functions by adding
or removing, as appropriate, those functions whose bounding spheres we
intersect.  The locations of these bounding spheres are likewise
precompiled into lists for efficient processing.  The memory overhead
due to this method is still much smaller than the storage requirements
for the actual function values.

\section{Results}\label{sec:results}
In this section we calculate common quantities using the localized
basis set on different systems. The results are compared to the
complete basis set limit, i.e.\ a well converged grid calculation.
Note that this comparison can be done in a very systematic way since
the calculations on the grid share the same approximations and mostly
the same implementation as the calculations performed with the
localized basis. All the results presented in this section have been
obtained using PAW setups from the extensive GPAW library, freely
available online\cite{setups}.

\subsection{Molecules}
\begin{figure*}
  \includegraphics[width=0.95\textwidth]{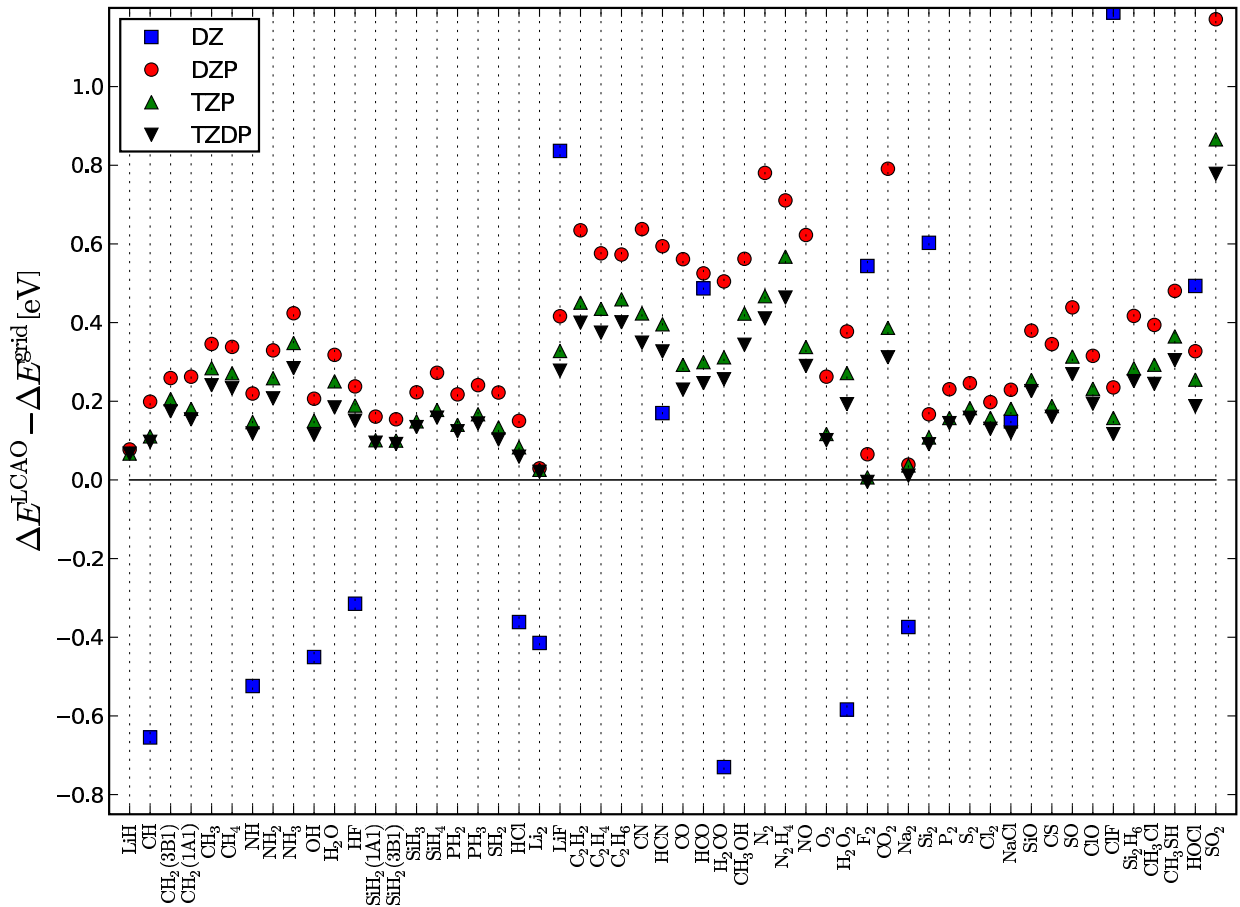}
  \caption{(Color online) PBE atomization energies from the G2-1 dataset, relative to the
    grid values. The corresponding Mean Absolute Errors with respect
    to the grid values are: 1.71 eV (20.4\%) for Double Zeta (DZ);
    0.36eV (4.45\%) for Double Zeta Polarized (DZP); 0.25 eV (3.02\%)
    for Triple Zeta Polarized (TZP)and 0.20 eV (2.44\%) for Triple
    Zeta Double Polarized (TZDP).}
  \label{fig:g2data}
\end{figure*}

In order to assess the accuracy of the LCAO implementation for small
molecules, the PBE\cite{pbe} atomization energies for the G2-1
data-set\cite{Curtiss:1997} are considered.  The atomic coordinates
are taken from MP2(full)/6-31G(d) optimized geometries.  The error
with respect to the grid results is shown in Fig.\ \ref{fig:g2data}
for different basis sets.  This error is defined as
\begin{align}
  \Delta E^{\mrm{LCAO}} - \Delta E^{\mrm{grid}} &=
  E^{\mrm{LCAO}}_{\mrm{mol}} - \sum_{\mrm{atoms}}
  E^{\mrm{LCAO}}_{\mrm{atoms}} \nonumber\\
  &\quad- \left( E^{\mrm{grid}}_{\mrm{mol}} - \sum_{\mrm{atoms}}
    E^{\mrm{grid}}_{\mrm{atoms}} \right)
\end{align}
The reference grid results are well converged calculations
in very good agreement with the VASP\cite{vasp} and
Gaussian\cite{gaussian} codes.

The figure shows that enlarging the basis set, i.e.\ including more
orbitals per valence electron, systematically improves the results
towards the grid energies.

It must be noted that some differences with respect to the grid
atomization energies still remain, even in the case of large basis
sets. This is mainly due to the two following reasons. Firstly, the
basis functions are generated from spin-paired calculations and hence
they do not explicitly account for possible spin-polarized orbitals.
This is in practice accounted for by using larger basis-sets in order
to include more degrees of freedom in the shape of the wavefunctions.
Secondly, isolated atoms are difficult to treat because of their long
ranged orbitals. Actual basis functions are, in fact, obtained from
atomic calculations with an artificial confining potential thus
resulting in more confined orbitals.

\subsection{Solids}
The equilibrium bulk properties have been calculated for several
crystals featuring different electronic structures: simple metals (Li,
Na, Al), semiconductors (AlP, Si, SiC), ionic solids (NaCl, LiF, MgO)
transition metals (Fe, Cu, Pt) as well as one insulator (C).  The
results are shown in Fig.\ \ref{fig:bulk}.  For comparison with
grid-based calculations, the bar plots show the deviations from
grid-based results for each basis set, while the precise numbers
are shown in each of the corresponding tables.
All the calculations were performed with the solids in their lowest
energy crystal structure, using the PBE functional for exchange and
correlation\cite{pbe}.  The quantities were computed using the relaxed
structures obtained with the default, unoptimized basis sets.  The
calculations were generally spin-paired, i.e.\ non-magnetic, with the
exception of Fe and the atomic calculations used to get cohesive
energies.
\begingroup
\squeezetable
\begin{figure*}[p]
  \begin{minipage}{1.0\textwidth}
    \includegraphics[width=0.5\textwidth]{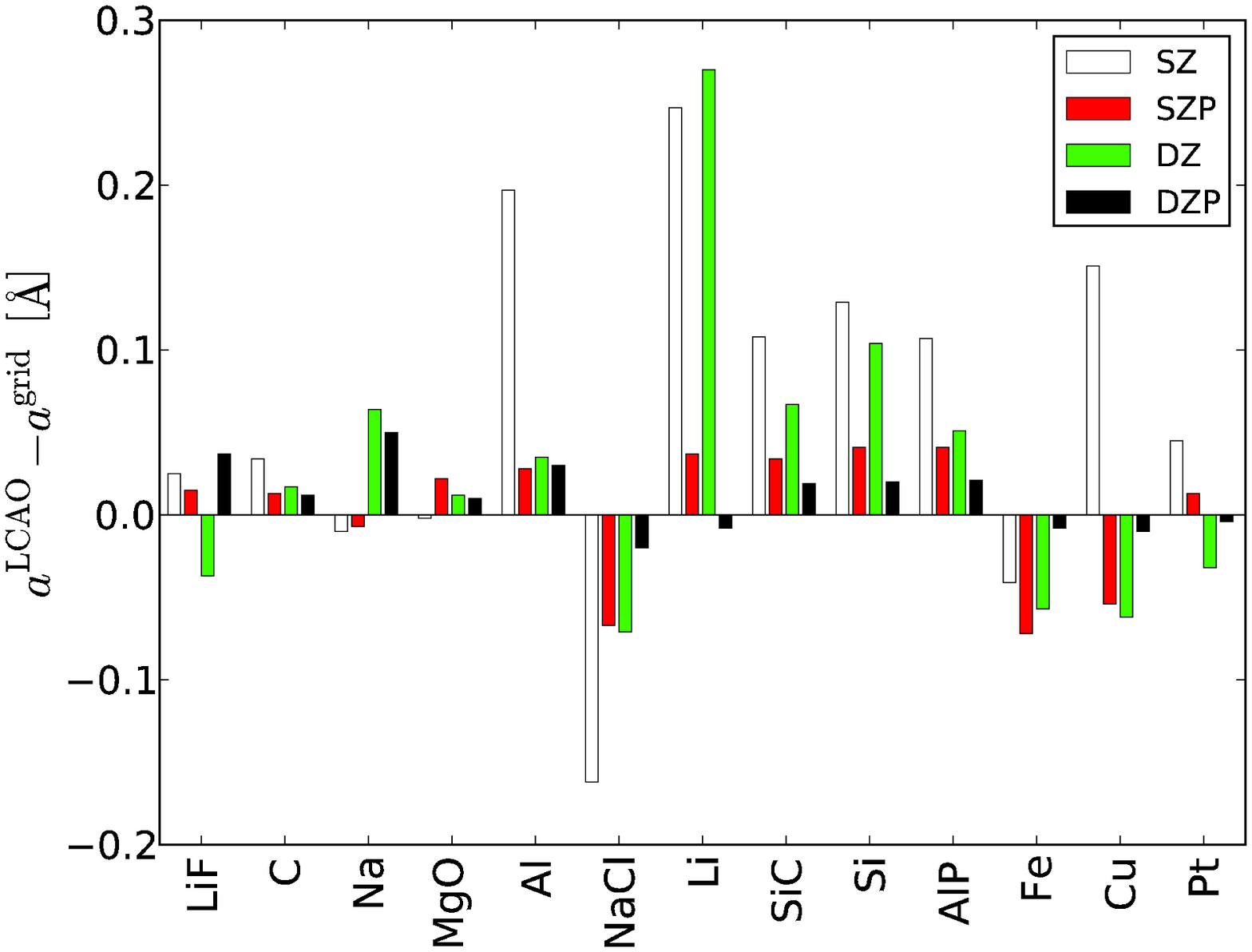}
    \label{fig:subfig1}
    \begin{tabular}[b]{llllll}
      \hline\hline
      &\multicolumn{5}{c}{$a$ (Å)}\\
      \cline{1-6}
      &     SZ&    SZP&     DZ&    DZP&   GRID\\
      \hline\hline
  LiF&  4.08&  4.08&  4.02&  4.10&  4.06 \\
    C&  3.61&  3.58&  3.59&  3.58&  3.57 \\
   Na&  4.18&  4.19&  4.26&  4.24&  4.19 \\
  MgO&  4.26&  4.28&  4.27&  4.27&  4.26 \\
   Al&  4.24&  4.07&  4.08&  4.07&  4.04 \\
 NaCl&  5.52&  5.62&  5.61&  5.67&  5.69 \\
   Li&  3.68&  3.47&  3.70&  3.43&  3.43 \\
  SiC&  4.50&  4.42&  4.46&  4.41&  4.39 \\
   Si&  5.60&  5.52&  5.58&  5.49&  5.48 \\
  AlP&  5.62&  5.55&  5.56&  5.53&  5.51 \\
   Fe&  2.80&  2.77&  2.78&  2.83&  2.84 \\
   Cu&  3.80&  3.59&  3.58&  3.64&  3.65 \\
   Pt&  4.02&  3.99&  3.95&  3.98&  3.98 \\
\hline
MAE &  0.097& 0.034& 0.068& 0.019 \\
MAE \% &  2.33& 0.84& 1.70& 0.45\\
\hline\hline\\
\end{tabular}
\includegraphics[width=0.5\textwidth]{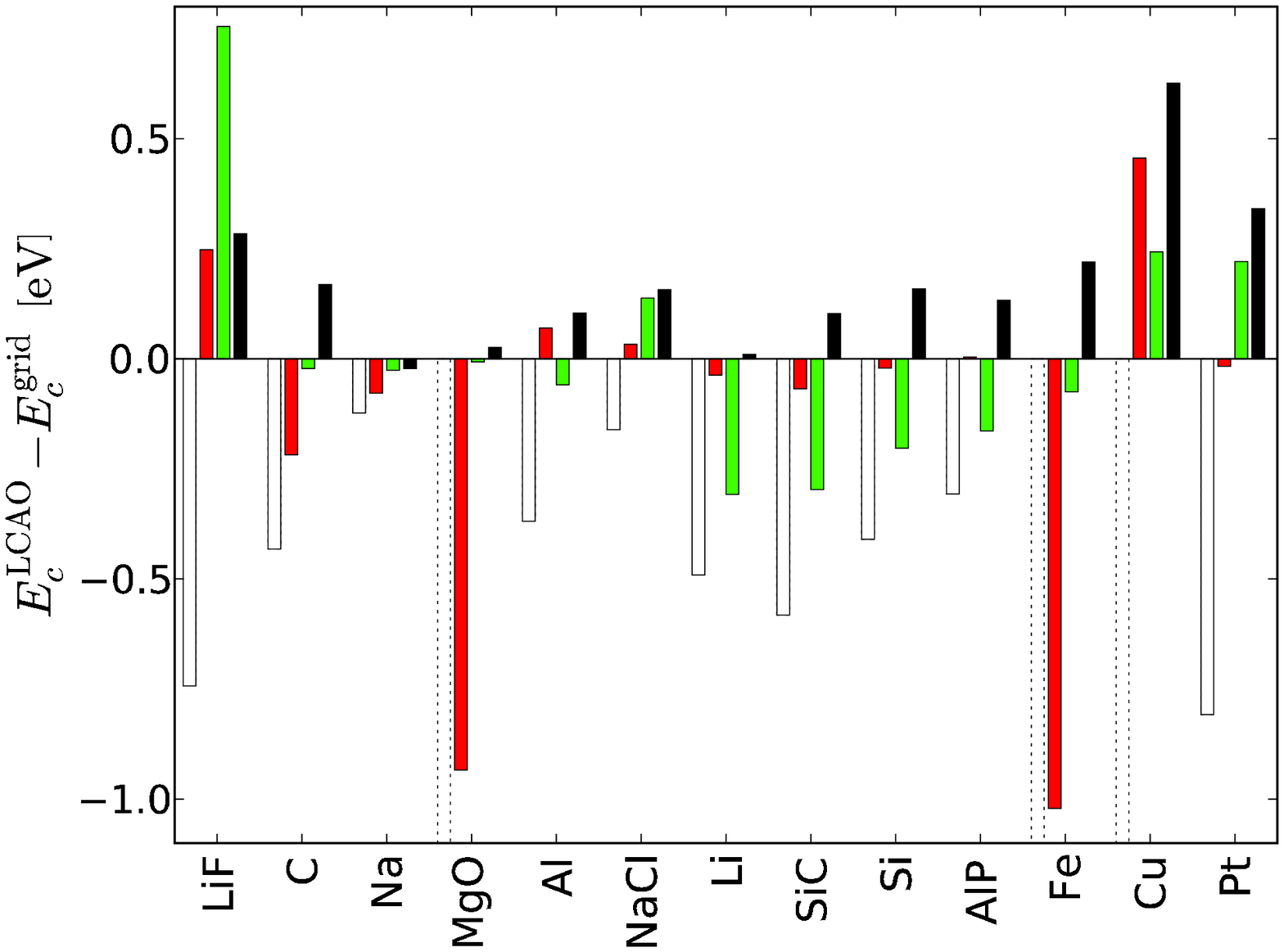}
\label{fig:subfig2}
\begin{tabular}[b]{llllll}
  \hline\hline
  &\multicolumn{5}{c}{$E_c$ (eV)}\\
  \cline{1-6}
  &     SZ&    SZP&     DZ&    DZP&   GRID\\
  \hline\hline
  
  LiF&  3.49&  4.48&  4.99&  4.52&  4.24 \\
    C&  7.29&  7.51&  7.70&  7.89&  7.72 \\
   Na&  0.97&  1.02&  1.07&  1.07&  1.09 \\
  MgO&  2.81&  4.01&  4.94&  4.97&  4.95 \\
   Al&  3.07&  3.51&  3.38&  3.54&  3.43 \\
 NaCl&  2.94&  3.14&  3.24&  3.26&  3.10 \\
   Li&  1.13&  1.58&  1.31&  1.63&  1.62 \\
  SiC&  5.80&  6.31&  6.08&  6.48&  6.38 \\
   Si&  4.14&  4.52&  4.34&  4.71&  4.55 \\
  AlP&  3.77&  4.09&  3.92&  4.21&  4.08 \\
   Fe&  1.34&  3.83&  4.77&  5.07&  4.85 \\
   Cu&  2.38&  3.97&  3.75&  4.14&  3.51 \\
   Pt&  4.54&  5.33&  5.57&  5.69&  5.35 \\
\hline
MAE &  0.86& 0.25& 0.19& 0.18 \\
MAE \% & 20.70& 5.86& 5.51& 4.40 \\
\hline\hline\\
\end{tabular}
\includegraphics[width=0.5\textwidth]{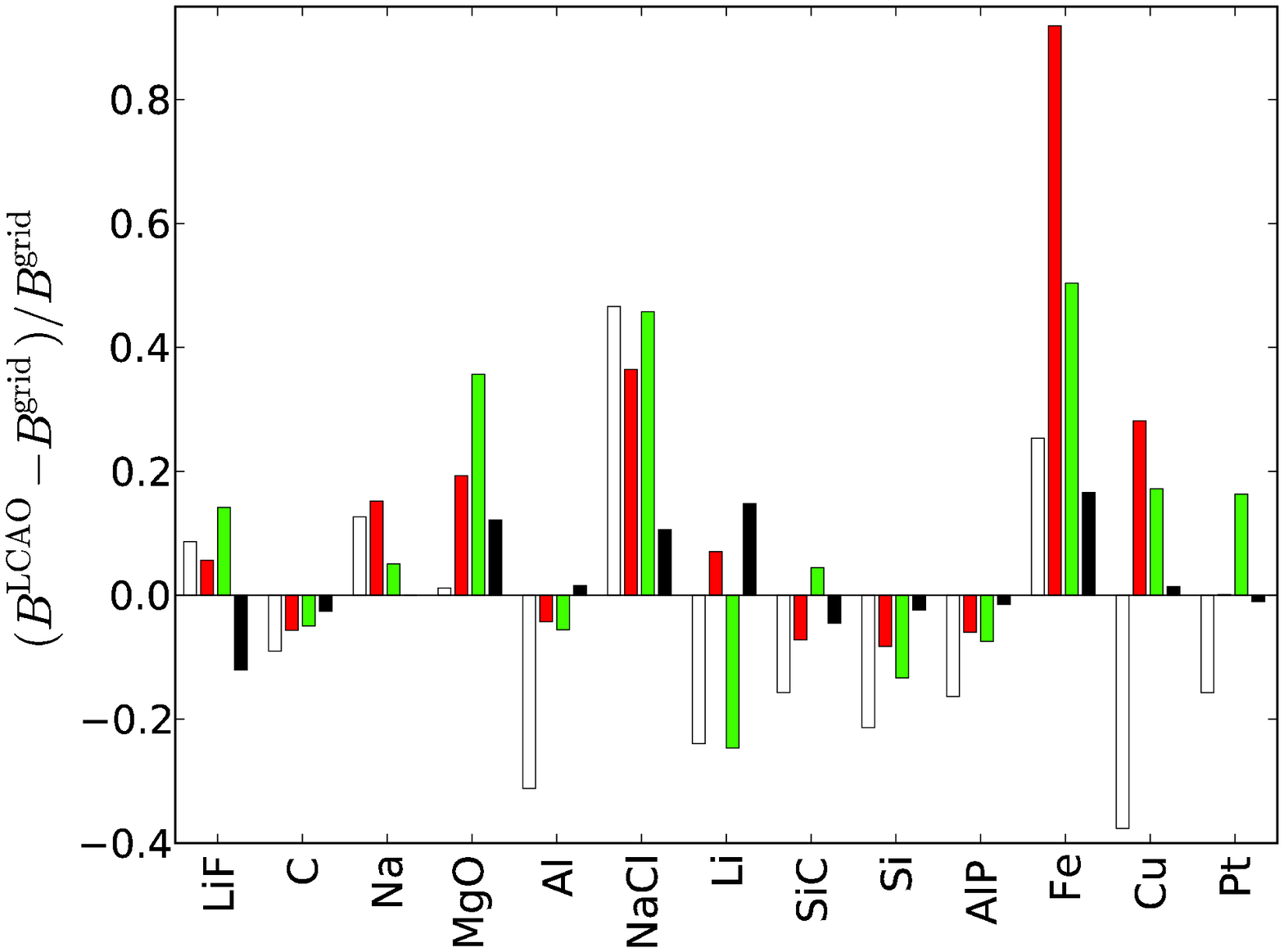}
\label{fig:subfig3}
\begin{tabular}[b]{llllll}
  \hline\hline
  &\multicolumn{5}{c}{$B$ (GPa)}\\
        \cline{1-6}
        &     SZ&    SZP&     DZ&    DZP&   GRID\\
        \hline\hline
  LiF&    87&    84&    91&    70&    80 \\
    C&   394&   408&   411&   422&   433 \\
   Na&   8.9&   9.1&   8.3&   7.9&   7.9 \\
  MgO&   156&   184&   209&   173&   154 \\
   Al&    53&    74&    73&    79&    77 \\
 NaCl&    35&    32&    34&    26&    24 \\
   Li&  10.8&  15.2&  10.7&  16.3&  14.2 \\
  SiC&   178&   196&   221&   202&   211 \\
   Si&    70&    81&    77&    86&    88 \\
  AlP&    69&    77&    76&    81&    82 \\
   Fe&   248&   379&   297&   231&   198 \\
   Cu&    88&   181&   166&   143&   141 \\
   Pt&   224&   266&   309&   263&   266 \\
\hline
MAE &  22.9& 24.8& 23.2&  7.4 \\
MAE \% &  20.4& 18.2& 18.8&  6.3 \\
\hline\hline
\end{tabular}
\end{minipage}

\caption{(Color online) Deviations in cohesive energy (top), lattice
  parameter (middle) and relative bulk modulus (bottom) from the
  converged grid results.  The largest bars have been truncated and
  are shown with dotted edges -- see the corresponding tables for the
  precise values.}
\label{fig:bulk}
\end{figure*}
\endgroup

The overall agreement with the real-space grid is excellent: about
0.5\% mean absolute error in the computation of lattice constants, 4\%
in cohesive energies and 5-8\% for bulk moduli using DZP basis sets.
Notice that in many cases remarkably good results can be obtained even
with a small SZP basis, particularly for lattice constants. This shows that structure optimizations with the LCAO code are
likely to yield very accurate geometries. This is probably due to the
fact that calculations of equilibrium structures only involve energy
differences between very similar structures, i.e. not with respect to
isolated atoms, thus leading to larger error cancellations.

With DZP the primary source of error in cohesive energy comes from the
free-atom calculation, where the confinement of each orbital raises
the energy levels by around 0.1 eV.  Thus, atomic energies are
systematically overestimated, leading to stronger binding.  This error
can be controlled by using larger basis set cutoffs, i.e.\ choosing
smaller orbital energy shifts during basis generation.

\subsection{Structure optimizations}
LCAO calculations tend to reproduce geometries of grid-based
calculations very accurately.  In structure optimizations, the LCAO
code can therefore be used to provide a high-quality initial guess for
a grid-based structure optimization.  

While it is trivial to reuse a geometry obtained in one code for a
more accurate optimization in another, our approach is practical
because the two representations share the exact same framework.  Thus
the procedure is seamless as well as numerically consistent, in the
sense that most of the operations are carried out using the same
approximations, finite-difference stencils and so on.  With
quasi-Newton methods, the estimate of the Hessian matrix generated
during the LCAO optimization can be reused as well.  For most
non-trivial systems, an LCAO calculation is between 25 and 30 times
faster than a grid calculation, making the cost of the LCAO optimization
negligible.

\begin{figure}
  \includegraphics[width=0.48\textwidth]{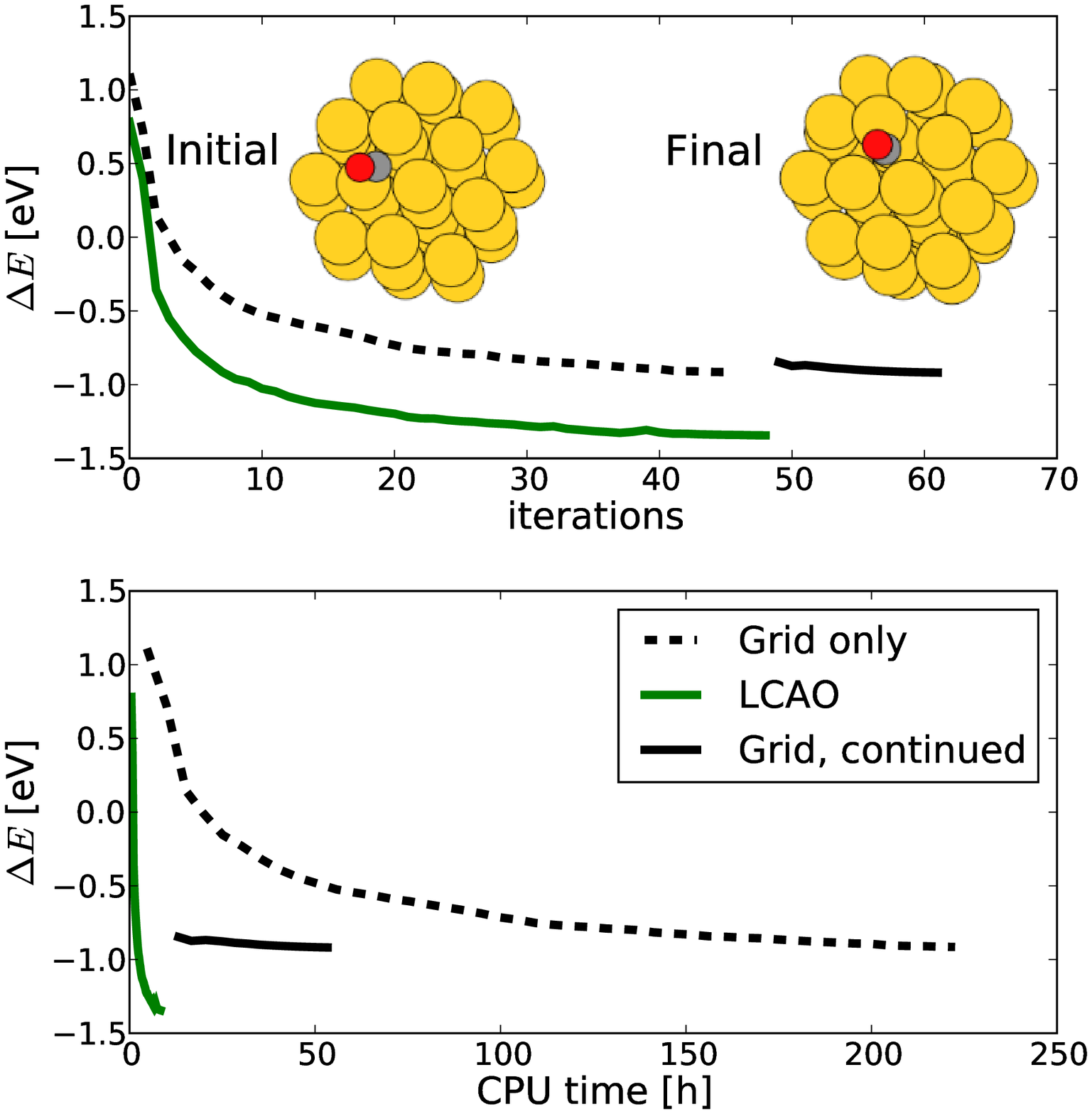}
  \caption{(Color online) The energy as a function of iteration count (top) as well
    as CPU time (bottom) in structure optimizations.  Shows a
    grid-based and an LCAO based structure optimization plus the
    continuation of the LCAO optimization after switching to the grid
    representation.}
  \label{fig:combinedrelaxation}
\end{figure}

Fig.\ \ref{fig:combinedrelaxation} shows a performance comparison when
reusing the positions and Hessian from a LCAO-based structure
optimization for a grid-based one, using the default basis set.  The
system is a 38-atom truncated octahedral gold cluster with CO
adsorbed, with the initial and final geometries shown in the inset.

A purely grid-based optimization takes 223 CPU hours while a purely
LCAO-based one, requiring roughly the same number of steps, takes 8.4
CPU hours.  A further grid-based optimization takes 45 CPU hours, for
a total speedup factor of 4.  The value of an initial LCAO
optimization is of course higher if the initial guess is worse.
For systems where a large fraction of the time is spent close to the
converged geometry, the speedup may not be as significant.

The energy reference corresponds to the separate cluster and molecule
at optimized geometries -- the total energy difference between an LCAO
and a grid calculation is otherwise around 30 eV.  It is therefore
important to choose an optimization algorithm which will handle such a
shift well.  The present plots use the L-BFGS 
algorithm\cite{sheppard,nocedal}
(limited memory Broyden-Fletcher-Goldfarb-Shanno)
 from the Atomic
Simulation Environment.\cite{ase}
\section{Conclusions}
We have described the implementation of a localized basis in the grid
based PAW code GPAW, and tested the method on a variety of molecules
and solids. The results for atomization energies, cohesive energies,
lattice parameters and bulk moduli were shown to converge towards the
grid results as the size of the LCAO basis was increased.  Structural
properties were found to be particularly accurate with the LCAO basis.
It has been demonstrated how the LCAO basis can be used to produce
accurate initial guesses (both for the electron wavefunctions, atomic
structure, and Hessian matrix) for subsequent grid-based calculations
to increase efficiency of high-accuracy grid calculations.

The combination of the grid-based and LCAO methods in one code
provides a flexible, simple and smooth way to switch between the two
representations. Furthermore the PAW formalism itself presents
significant advantages: it is an all-electron method, which eliminates
pseudopotential errors, and it allows the use of coarser grids than
norm-conserving pseudopotentials, which increases efficiency.

Finally, the LCAO method enables GPAW to perform calculations
involving Green's function, which intrinsically need a basis set with
finite support. Current developments along these lines include electron transport
calculations, electron-phonon coupling and STM simulations.

\begin{acknowledgments}
  The authors acknowledge support from the Danish Center for
  Scientific Computing through grant HDW-1103-06. The Center for
  Atomic-scale Materials Design is sponsored by the Lundbeck
  Foundation.
\end{acknowledgments}
\appendix
\section{Force formula}\label{appendixforces}
The force on atom $a$ is found by taking the derivative of the total
energy with respect to the atomic position $\ve R^a$.  We shall use
the chain rule on Eq.\ \eqref{totalenergy}, taking $\rho_{\mu\nu}$,
$D_{ij}^a$, $\ps n(\ve r)$, $\ps \rho(\ve r)$, $T_{\mu\nu}$ and $\bar
v(\ve r)$ to be \emph{separate variables} for the purposes of partial
derivatives:
\begin{align}
  \pdiff{E}{\ve R^a} &= 
  \sum_{\mu\nu}\pdiff{E}{\rho_{\nu\mu}}\pdiff{\rho_{\nu\mu}}{\ve R^a}
  + \sum_{bij} \pdiff{E}{D_{ji}^b} \pdiff{D_{ji}^b}{\ve R^a} \nn\\
  &\quad+ \int \fdiff{E}{\ps n(\ve r)} \pdiff{\ps n(\ve r)}{\ve R^a} \idr
  + \int \fdiff{E}{\ps \rho(\ve r)} \pdiff{\ps \rho(\ve r)}{\ve R^a} \idr
  \nn\\&\quad
  + \sum_{\mu\nu}\pdiff{E}{T_{\mu\nu}} \pdiff{T_{\mu\nu}}{\ve R^a}
  + \int \fdiff{E}{\bar v(\ve r)} 
  \pdiff{\bar v(\ve r)}{\ve R^a} \idr,
  \label{chainrule}
\end{align}
where $\bar v(\ve r) = \sum_a \bar v^a(|\ve r - \ve R^a|)$.  The
remaining quantities in the energy expression pertain to isolated
atoms, and thus do not depend on atomic positions.  The first term of
Eq.\ \eqref{chainrule} is
\begin{align}
  \sum_{\mu\nu}\pdiff{E}{\rho_{\nu\mu}} \pdiff{\rho_{\nu\mu}}{\ve R^a}
  &=
  2\Re \sum_{\mu\nu n} H_{\mu\nu} c_{\nu n} f_n \pdiff{c_{\mu n}^*}{\ve R^a}
  \nn\\
  &= 2\Re \sum_{\mu\nu n} 
  \pdiff{c_{\mu n}^*}{\ve R^a} S_{\mu\nu} c_{\nu n} \epsilon_n  f_n
  ,
  \label{dEdrhodrhodR1}
\end{align}
where we have used Eqs.\ \eqref{densitymatrix} and
\eqref{hamiltonianmatrix} in the first step, and Eq.\
\eqref{eigenvalueproblem} in the second.  When the atoms are displaced
(infinitesimally), the coefficients must change to accommodate the
orthogonality criterion.  This can be incorporated by requiring the
derivatives of each side of Eq.\ \eqref{orthogonalitycondition} to be
equal, implying the relationship
\begin{align}
  - \sum_{\mu\nu} c_{\mu n}^* \pdiff{S_{\mu\nu}}{\ve R^a} c_{\nu n}
  = 2\Re \sum_{\mu\nu} 
    \pdiff{c_{\mu n}^*}{\ve R^a} S_{\mu\nu} c_{\nu n}.
\end{align}
Inserting this into Eq.\ \eqref{dEdrhodrhodR1} yields
\begin{align}
  \sum_{\mu\nu}\pdiff{E}{\rho_{\nu\mu}} \pdiff{\rho_{\nu\mu}}{\ve R^a}
  &= - \sum_{\mu\nu n} \pdiff{S_{\mu\nu}}{\ve R^a} c_{\nu n} 
  \epsilon_n f_n c_{\mu n}^* \nn\\
  &= -\sum_{\mu\nu} \pdiff{S_{\mu\nu}}{\ve R^a} E_{\nu \mu},
  \label{dEdrhodrhodR2}
\end{align}
where we have introduced the matrix
\begin{align}
  E_{\nu\mu} = \sum_n c_{\nu n} \epsilon_n f_n c_{\mu n}^*
  = \sum_{\lambda \xi} S^{-1}_{\nu\lambda} H_{\lambda\xi} \rho_{\xi\mu}.
\end{align}
The equivalence of these forms follows from Eq.\
\eqref{eigenvalueproblem}.  The overlap matrix elements $S_{\mu\nu}$
depend on $\ve R^a$ through the two-center integrals $\Theta_{\mu\nu}$
and $P_{i\mu}^b$.  The derivative of a two-center integral can be
nonzero only if exactly one of the two involved atoms is $a$, and for
nonzero derivatives, the sign changes if the indices are swapped.
Taking these issues into account, Eq.\ \eqref{dEdrhodrhodR2} is split
into those three terms in Eq.\ $\eqref{forces}$ which contain $E_{\nu\mu}$.

In the second term in Eq.\ \eqref{chainrule}, we take the
$D_{ij}^b$-dependent derivative for fixed $\rho_{\nu\mu}$, which by
Eq.\ \eqref{DeltaH} evaluates to
\begin{align}
  \sum_{bij}\pdiff{E}{D_{ji}^a} \pdiff{D_{ji}^a}{\ve R^a}
  = 2 \Re \sum_{bij\mu\nu} 
  P_{i\mu}^{b*} \Delta H_{ij}^b \pdiff{P_{j\nu}^b}{\ve R^a} \rho_{\nu\mu}.
\end{align}
Again most of the two-center integral derivatives are zero.  A
complete reduction yields the two terms in Eq.\ \eqref{forces} which
depend on the $\ve A_{\mu\nu}^b$ vectors.

Using Eq. \eqref{lcaopseudodensity}, the third term of Eq.\
\eqref{chainrule} is
\begin{align}
  \int \fdiff{E}{\ps n(\ve r)} \pdiff{\ps n(\ve r)}{\ve R^a} \idr
  &= \int \ps v(\ve r) \pdiff{\ps n(\ve r)}{\ve R^a} \idr \nn\\
  &= 2 \Re \sum_{\mu\nu} \left[ \int 
    \pdiff{\Phi_\mu^*(\ve r)}{\ve R^a} \ps v(\ve r) \Phi_\nu(\ve r)
  \right] \rho_{\nu\mu} \nn\\
  &\quad + \int \ps v(\ve r) \pdiff{\ps n_c^a(|\ve r - \ve R^a|)}{\ve R^a} \idr.
\end{align}
The sum over $\mu$ can be restricted to $\mu \in a$.

Consider the fourth term of Eq. \eqref{chainrule}.  Aside from $\ps
n(\ve r)$ and $D_{ij}^b$, which are considered fixed as per the chain
rule, the pseudo charge density $\ps \rho(\ve r)$ depends only on the
locations of the compensation charge expansion functions $\ps
g_L^a(\ve r)$ which move rigidly with the atom, so
\begin{align}
  \int \fdiff{E}{\ps \rho(\ve r)} \pdiff{\ps \rho(\ve r)}{\ve R^a} \idr 
  &= \int \ps v_H(\ve r) \fdiff{\ps \rho(\ve r)}{\ps Z(\ve r)}
  \sum_{bL} \fdiff{\ps Z(\ve r)}{\ps g_L^b(\ve r)} \pdiff{\ps
    g_L^b(\ve r)}{\ve R^a} \idr \nn\\
  &= \int \ps v_H(\ve r) \sum_L Q_L^a \pdiff{\ps g_L^a(\ve r)}{\ve R^a} \idr.
\end{align}
The kinetic term from Eq.\ \eqref{chainrule} is
\begin{align}
  \sum_{\mu\nu} \pdiff{E}{T_{\mu\nu}}\pdiff{T_{\mu\nu}}{\ve R^a} 
  = \sum_{\mu\nu} \pdiff{T_{\mu\nu}}{\ve R^a} \rho_{\nu\mu}
\end{align}
and can also be restricted to $\mu \in a$.  Finally, the contribution
from the local potential $\bar v^a(\ve r)$ is simply
\begin{align}
  \int \fdiff{E}{\bar v(\ve r)} \pdiff{\bar v(\ve r)}{\ve R^a} \idr
  = \int \ps n(\ve r) \pdiff{\bar v^a(\ve r - \ve R^a)}{\ve R^a} \idr.
\end{align}
By now we have considered all position-dependent variables in the
energy expression, and have obtained expressions for all terms present
in Eq.\ \eqref{forces}.
\bibliographystyle{apsrev}

\begin{thebibliography}{23}
\expandafter\ifx\csname natexlab\endcsname\relax\def\natexlab#1{#1}\fi
\expandafter\ifx\csname bibnamefont\endcsname\relax
  \def\bibnamefont#1{#1}\fi
\expandafter\ifx\csname bibfnamefont\endcsname\relax
  \def\bibfnamefont#1{#1}\fi
\expandafter\ifx\csname citenamefont\endcsname\relax
  \def\citenamefont#1{#1}\fi
\expandafter\ifx\csname url\endcsname\relax
  \def\url#1{\texttt{#1}}\fi
\expandafter\ifx\csname urlprefix\endcsname\relax\def\urlprefix{URL }\fi
\providecommand{\bibinfo}[2]{#2}
\providecommand{\eprint}[2][]{\url{#2}}

\bibitem{hohenbergkohn}
  P.\ Hohenberg, W.\ Kohn, Phys.\ Rev.\ \textbf{136} B664 (1964).

\bibitem{kohnsham}
  W.\ Kohn, L.\ J.\ Sham, Phys.\ Rev.\ \textbf{140} B1133 (1965).

\bibitem{payne_RMP}
M. C. Payne, M. P. Teter, D. C. Allan, T. A. Arias, and J. D. Joannopoulos,
Rev. Mod. Phys. {\bf 64}, 1045 (1992).

\bibitem{kleinmanbylander82}
L. Kleinman and D. M. Bylander, Phys. Rev. Lett. {\bf 48}, 1425 (1982).

\bibitem{vanderbilt}
  D.\ Vanderbilt, Phys.\ Rev.\ B \textbf{14}, 7892 (1990).

\bibitem{hgh}
  C. Hartwigsen, S. Goedecker, and J. Hutter, Phys.\ Rev.\ B \textbf{58}, 3641 (1998).

\bibitem{Blochl:1994}
  P.\ E.\ Bl{\"o}chl,
  Phys. Rev. B {\bf 50}, 17953 (1994).

\bibitem{vasp}
  G.\ Kresse, J.\ Hafner, 
  Phys.\ Rev.\ B, \textbf{47}, 558, (1993).

\bibitem{dacapo}
B. Hammer, L.B. Hansen, and J.K. N{\o}rskov,
  Phys.\ Rev.\ B {\bf 59}, 7413 (1999).

\bibitem{goedecker}
L. Genovese, A. Neelov, S. Goedecker, T. Deutsch, A. Ghasemi,
O. Zilberberg, A. Bergman, M. Rayson, R Schneider, , J. Chem. Phys. {\bf 129}, 014109 (2008).

\bibitem{arias}
T. A. Arias, Rev. Mod. Phys. {\bf 71}, 267 (1999).

\bibitem{bernholc96}
E. L. Briggs, D. J. Sullivan, and J. Bernholc, 
Phys. Rev. B {\bf 54}, 14362 (1996).

\bibitem{Mortensen:2005}
  J.\ J.\ Mortensen, L.\ B.\ Hansen, and K.\ W.\ Jacobsen
  Phys. Rev. B {\bf 71}, 035109 (2005).

\bibitem{gaussian}
  Gaussian 03, Revision C.02, 
  Gaussian, Inc., Wallingford CT, (2004).

\bibitem{Sankey:1989}
  O.\ F.\ Sankey and D.\ J.\ Niklewski
  Phys. Rev. B {\bf 40}, 3979 (1989).

\bibitem{siesta}
  J.\ M.\ Soler, E.\ Artacho, J.\ D.\ Gale, A.\ Garc\'{i}a, J.\ Junquera,
P.\ Ordej\'{o}n, D.\ Sanchez-Portal
  J. Phys. Cond. Matter {\bf 14}, 2745-2779 (2002).

\bibitem{siesta2}
  J.\ Junquera, O.\ Paz, D.\ Sanchez-Portal, E.\ Artacho,
  Phys.\ Rev.\ B \textbf{64}, 235111 (2001).

\bibitem{setups}
  http://wiki.fysik.dtu.dk/gpaw/setups/setups.html

\bibitem{Curtiss:1997} 
  L.\ A.\ Curtiss, Krishnan Raghavachari, P.\ C.\ Redfern, J.\ A.\ Pople,
  J. Chem. Phys. {\bf 106}, 1063 (1997).

\bibitem{ase}
  S.\ Bahn and K.\ W.\ Jacobsen, Comput. Sci. Eng. 4, 56 (2002).

\bibitem{pbe}
  J.\ P.\ Perdew, K.\ Burke, M.\ Ernzerhof, Phys.\ Rev.\ Lett.\ \textbf{77},
  3865 (1996).

\bibitem{sheppard}
  D.\ Sheppard, R.\ Terrell, and G.\ Henkelman, J.\ Chem.\ Phys.\ \textbf{128},
  134106 (2008).

\bibitem{nocedal}
  J.\ Nocedal, Math.\ Comput.\ \textbf{35}, 773 (1980).

\end{thebibliography}

\end{document}